\documentclass[11pt]{article}

\usepackage{amsmath,amssymb}
\usepackage{slashed}
\usepackage{xcolor} 
\usepackage{graphicx}
\usepackage{url}

\newcommand{\be}{\begin{equation}}
\newcommand{\ee}{\end{equation}}
\newcommand{\bea}{\begin{eqnarray}}
\newcommand{\eea}{\end{eqnarray}}
\def\Re{{\cal R \mskip-4mu \lower.1ex \hbox{\it e}\,}}
\def\Im{{\cal I \mskip-5mu \lower.1ex \hbox{\it m}\,}}

\def\tev{\,{\ifmmode\mathrm {TeV}\else TeV\fi}}
\def\gev{\,{\ifmmode\mathrm {GeV}\else GeV\fi}}
\def\mev{\,{\ifmmode\mathrm {MeV}\else MeV\fi}}

\begin{document}

\begin{center}
\vspace*{15mm}

\vspace{1cm}
{\Large \bf  Probing the top quark flavour-changing neutral current at a future electron-positron collider} \\
\vspace{1cm}

{\bf  Hoda Hesari${}^1$, Hamzeh Khanpour${}^{1,2}$, Morteza Khatiri Yanehsari${}^{1,3}$, \\
 Mojtaba Mohammadi Najafabadi${}^1$ }

 \vspace*{.5cm}
{\small\sl$^1\,$School of Particles and Accelerators, Institute for Research in Fundamental Sciences (IPM) P.O. Box 19395-5531, Tehran, Iran } \\
{\small\sl$^2\,$Department of Physics, Mazandaran University of Science and Technology, P.O. Box 48518-78413, Behshahr, Iran } \\
{\small\sl$^3\,$Department of Physics, Ferdowsi University of Mashhad, P.O.Box 1436, Mashhad, Iran }

\vspace*{.2cm}

\end{center}

\vspace*{10mm}

%
%

\begin{abstract}
We present a study to examine the sensitivity of a future $e^{-} e^{+}$ collider
to the anomalous top flavour-changing neutral current (FCNC) to the gluon.
To separate signal from background a multivariate analysis is performed on top-quark pair and background
events, where one top quark is considered to follow the dominant Standard Model (SM) decay, $t \rightarrow Wb$,
and the other top decays through FCNC, $t\rightarrow qg$, where $q$ is a $u-$ or a $c-$quark.
The analysis of fully-hadronic FCNC decay of the $t \bar{t}$ pair is also presented.
The $95\%$ confidence level limits on the top-quark anomalous couplings are obtained for different values
of the center-of-mass energies and integrated luminosities.
\end{abstract}

\vspace*{3mm}



%
%
\section{Introduction}\label{Introduction}

The top-quark, which is the heaviest known elementary particle up to now,
plays a special role in search for new physics beyond the Standard Model (SM) in particular through precise measurement
of its couplings with other particles.
The large mass of the top quark, $M_{top} = 173.34 \pm 0.27 (stat) \pm 0.71 (syst)$~\cite{ATLAS:2014wva},
that is close to the scale of electroweak symmetry breaking and its interactions with other particles such as
the Higgs boson make it an excellent object to investigate the validity of the SM.
The anomalous interactions of the top quark can occur in various flavour-changing neutral current (FCNC) processes
like $t \rightarrow q X$, where $X = g, \gamma, Z$ or Higgs.
In \cite{sv1}, the anomalous $tq\gamma$ and $tqZ$ have been probed
at a future electron-positron collider.
In the present study, we focus on the top quark FCNC interactions involving
the top-quark, a light quark $q$, ($u$- or $c$-quark), and a gluon.
In the SM, the FCNC transition of $t\rightarrow qg$ ($q=u, c)$
is forbidden at tree level due to the Glashow-Iliopoulos-Mainai (GIM) mechanism \cite{gim} and only can proceed through the loop corrections.
In the SM framework, the loop-level
branching ratio for  $t\rightarrow qg$ ($q = u, c)$ is of the order of $10^{-12}$ ~\cite{th1,th2}.
Clearly, a lot of data is needed to enable us to observe such a decay process and measure this small branching ratio.
Various models beyond the SM could lead to a very large increase in FCNC processes involving the top-quark.
Thus, any evidence of such processes will indicate the existence of new physics.
In models beyond SM such as MSSM, Technicolor, extra dimensions models higher branching ratios up to $10^{-3}$ - $10^{-5}$
are predicted \cite{bsm1,bsm2,bsm3} which can be tested by present high energy experiments.
There are several phenomenological studies in search for the anomalous $tqg$ couplings at the Tevatron and LHC
and other experiments through different channels \cite{tqg1,tqg2,tqg3,Etesami:2010ig}.
At present the best and up-to-date experimental limits on the $tqg$ branching fractions come from
the direct top production process at the Large Hadron Collider (LHC) by the
ATLAS Collaboration, $ Br(t \rightarrow ug) < 3.1 \times 10^{-5}$ and
$Br(t \rightarrow cg) < 1.6 \times 10^{-4}$ at a center-of-mass energy of $\sqrt{s} = 8$ TeV
corresponding to an integrated luminosity of ${\cal L}_{int} = 14.2$ fb$^{-1}$~\cite{TheATLAScollaboration:2013vha,Yazgan:2013pxa}.

It is expected that the future TeV scale linear colliders such as Compact Linear Collider (CLIC) or 
International Linear Collider (ILC) would complete the LHC
probes and even in some processes can improve the measurements and limits.
The high luminosity and clean experimental environments of the TeV scale $e^{-} e^{+}$ collider
make it an excellent precision machine
for the investigation of the top-quark properties. It also provides us an
important opportunity for precise measurements of the FCNC couplings in top quark sector \cite{sv1,sv2}.
For example, in \cite{sv1,sv2} it has been shown that the branching ratios of the top quark
decay into a photon and a Z-boson can be measured up to the order of $10^{-6}$ at a linear electron-positron collider.

The $e^{-}e^{+}$ collider, CLIC, is designed to operate with the center-of-mass energies
of $\sqrt{s} = 0.5$, $1.5$ and $3$ TeV corresponding to total luminosity
of $L = 2.3$, $3.2$ and $5.9 \times 10 ^{34}$ cm$^{-2}$ s$^{-1}$,
respectively~\cite{Lebrun:2012hj,Miyamoto:1425915,Brau:2012hv,Aicheler:2012bya}.
The design of ILC is to work at the center-of-mass energies from  
$\sqrt{s} = 0.25$ TeV to  0.5 TeV with the option of upgrading to 1 TeV. 
The plan for the ILC instantaneous luminosities is to reach $10^{33}-10^{34}$ cm$^{-2}$ s$^{-1}$ \cite{ilc1,ilc2}.
One of the main differences
between the ILC and CLIC is the difference in the luminosity
spectrum (LS) of the these machines. The ILC luminosity spectrum has
a narrower peak of luminosity. This leads to an increase of the total luminosity
and consequently reducing the statistical uncertainty in the measurements.

In this work, we study the sensitivity of a future electron-positron collider (CLIC or ILC)
to the anomalous top flavour-changing neutral current (FCNC) to the gluon, $t-q-g$.
To separate signal from backgrounds, a multivariate technique is used.
We consider $e^{-}e^{+} \rightarrow t\bar{t} \rightarrow qg\ell^{+}\nu_{\ell}b(\bar{q}g\ell^{-}\bar{\nu}_{\ell}\bar{b})$ (semi-leptonic) and
$e^{-}e^{+} \rightarrow t\bar{t}\rightarrow q\bar{q}gg$ (full-hadronic) separately to search for the
anomalous FCNC interactions in  $t-q-g$ vertex.
The analysis can also be done in the full hadronic case with one of the top
quarks decays into $t\rightarrow W b \rightarrow jjj$ and the other top decays
through anomalous couplings. Because of the large background contribution
the results would be better than the semi-leptonic case therefore we do not
perform the analysis for this decay mode.
We consider the center-of-mass energies of $0.5$, $1$ and $1.5$ TeV, and for these
energies we analyses two cases, semi-leptonic and fully-hadronic decays of top-quark.

The presented paper is organised as follows:
In section~\ref{formalism}, we introduce the theoretical formalism which describes the FCNC processes.
Section~\ref{semi-leptonic} provides a full detailed description of the semi-leptonic channel in search for the
$tqg$ FCNC. The event selection and the methods of event classification into signal- and background-like events
using a multivariate analysis are also discussed in this section.
Our fully-hadronic analysis is presented in Sec.~\ref{Fully-hadronic}.
The results of the investigated FCNC processes, including expected sensitivities on the anomalous couplings
and corresponding branching fractions, are given in Sec.~\ref{results}. Discussions on 
some detector effects and systematic uncertainties are also presented in Sec.~\ref{results}.
Section~\ref{Conclusions} contains a summary and conclusions of the analysis.

%
%
\section{ Theoretical formalism }\label{formalism}

\begin{figure}[tbh]
\vspace{0.75cm}
\centering
\includegraphics[width=0.45\textwidth]{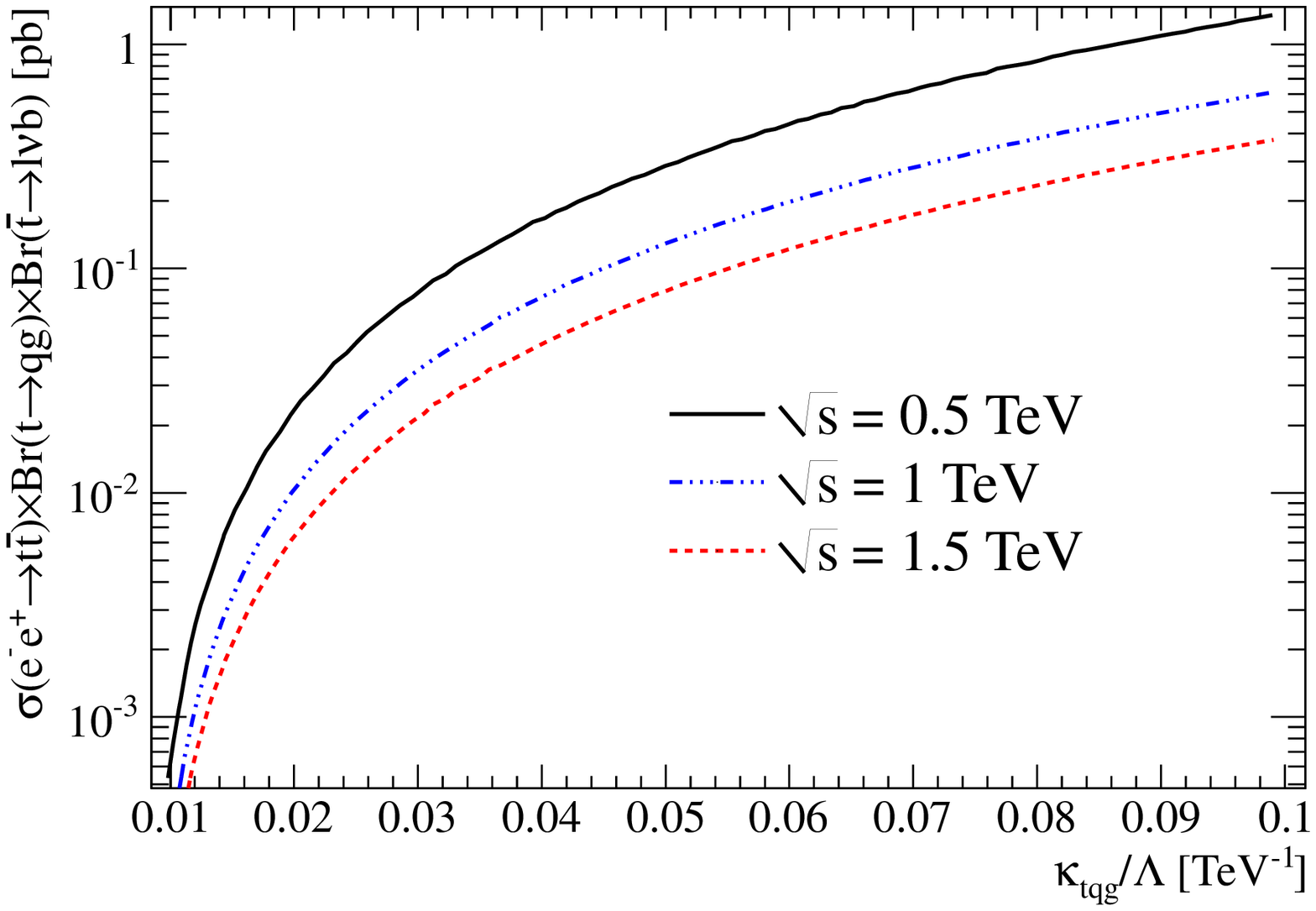}
\includegraphics[width=0.45\textwidth]{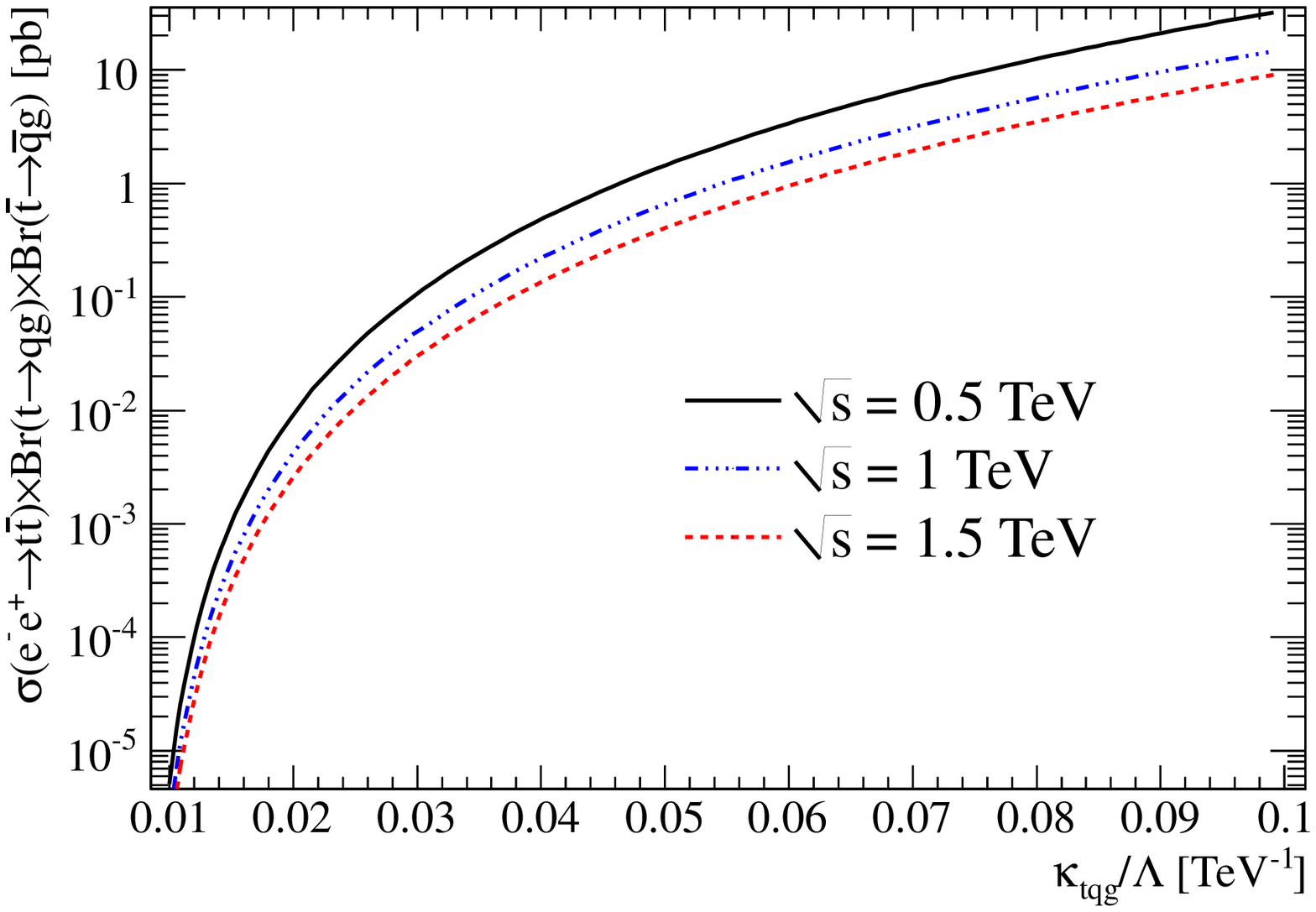}
\caption{The cross section times branching ratio of one (left) and both (right) of the top quarks decay anomalously into $q+g$ as a function
 of the anomalous coupling $\kappa_{tug}/\Lambda$ for $ \sqrt{s} = 0.5 $, $1$ and $1.5$ TeV.}
\label{Sigma-Ku}
\end{figure}

In this section, we give a brief overview of the theoretical framework for top FCNC which this analysis is based on.
In this work, to describe the FCNC couplings amongst the top quark, a light quark and a gluon ($tqg$) an effective Lagrangian approach is used.
The FCNC anomalous interaction in the vertex of $tqg$ can be written as follows ~\cite{Buchmuller:1985jz,lag1,lag2,lag3,Gao:2011fx}:
\begin{equation}\label{Effective-Lagrangian}
 {\cal L}_{eff} = \sum_{q=u,c} \frac{1}{ \Lambda } \, g_s\, \kappa_{tqg} \, \bar t \, \sigma^{\mu \nu} \, T^a \, \chi \, q \, G^a_{\mu\nu} + h.c \, .
\end{equation}
where the $\kappa_{tqg}$, with $q = u, c$, are dimensionless real parameters
that presents the strength of the anomalous couplings and strong coupling constant is denoted by $g_s$.
In Eq.\ref{Effective-Lagrangian},  $T^a = \frac{\lambda^{a}}{2}$ where $\lambda^{a}$ are the Gell-Mann matrices,
 $\Lambda$ is the new physics scale,
 $G^a_{\mu\nu}$  is the gluon field tensor and $\sigma ^{\mu \nu  }=\frac{i}{2}[\gamma ^{\mu },\gamma ^{\nu}]$.
In the effective Lagrangian $\chi = f_q^L \, P_L + f_q^R \, P_R$ with $P_L(P_R)$ operators
perform the left- (right-) handed projection and $f_q^{(L,R)}$ are chiral
parameters normalized to $|f_{q}^{L}|^{2} + |f_{q}^{R}|^{2} = 1$.
In Fig.~\ref{Sigma-Ku}, we show the top pair production cross section times the
branching ratios of one top decay anomalously into $q + g$ and another one decay leptonically (electron and muon)
as well as the top pair production cross section times the
branching ratios of both tops decay anomalously into $q + g$.
It is presented for different center-of-mass energies, $ \sqrt{s} = 0.5 $, $1$ and $1.5$ TeV versus
the anomalous coupling $\kappa_{tqg}/\Lambda$.
As it can be seen the $\sigma(e^{-}e^{+}\rightarrow t\bar{t}\rightarrow qg\ell^{-}\nu \bar{b} (\bar{q}g\ell^{+}\nu b))(\frac{\kappa_{tqg}}{\Lambda}=0.02~TeV^{-1})=22.2$ fb
and $\sigma(e^{-}e^{+}\rightarrow t\bar{t}\rightarrow qg\bar{q}g ( \bar{q}gqg))(\frac{\kappa_{tqg}}{\Lambda}=0.02~TeV^{-1})=9.6$ fb
for the center-of-mass energy of 0.5 TeV.
In order to calculate the cross section and simulate the events for the analysis, the FCNC effective Lagrangian
has been implemented in the { \sc FeynRules} package \cite{Christensen:2008py,Duhr:2011se} then
 the model has been imported to a Universal FeynRules Output (UFO) module \cite{Degrande:2011ua}
and finally inserted to the {\sc MadGraph} 5 ~\cite{Alwall:2011uj}. The values of the cross sections are found to be in agreement
with { \sc CompHEP} package \cite{Boos:2009un,Boos:2004kh}.

%
%
%
%

%
%
\section{Semi-leptonic channel}\label{semi-leptonic}

\begin{figure}[tbh]
\vspace{0.75cm}
\centerline{\includegraphics[width=0.50\textwidth]{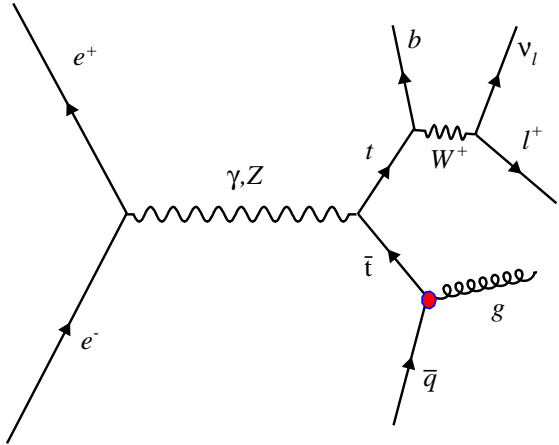}}
\caption{The representative Feynman diagram for the signal process in the semi-leptonic channel.}
\label{feynlep}
\end{figure}

This section presents the analysis of our signal and the corresponding backgrounds of semi-leptonic channel of $t \bar{t}$
events at the $e^{-} e^{+}$ collider.
In this channel, one of the top-quarks decays through SM decay mode of $t \rightarrow Wb \rightarrow \ell \nu_{\ell}b$  ($\ell = e, \mu$)
and the other one is considered to decay through FCNC into $q+g$, where $q$ is $u$ or $c$ quark.
The hadronic final states of W boson have larger background contribution which would not lead
to better sensitivity with respect to the semi-leptonic channel. 
Therefore, the leptonic decay modes of the W-boson that provide cleaner signature is considered.
The final state signal topology consists of an energetic lepton, neutrino (appears as missing momentum) and three
hadronic jets. One of the jets is originated from a b-quark.
The representative Feynman diagram for the signal process is depicted in Fig.~\ref{feynlep}.

Based on the expected signature of the signal events,
the background topology is therefore given by $W^{\pm} j j j \rightarrow  \ell^{\pm} \bar{\nu }_{\ell} j j j$.
In order to investigate the possibility of separating signal from the background events,
we use Monte Carlo (MC) simulation.
The MC generation of the signal sample, $ t \bar{t} \rightarrow b \ell \nu _{\ell} q g $ is done with
{ \sc CompHEP} ~\cite{Boos:2009un,Boos:2004kh} and the complete set of  $\ell \nu + 3 j$ background including the
 SM process $ t \overline{t} \rightarrow \ell \overline{\nu }_{\ell} j j j $ are generated
using { \sc MadGraph} ~\cite{Alwall:2011uj}. The symbol $j$ represents any jet that
originates from quarks and gluon.

To account for the resolution of detectors, we apply energy smearing effects to the
final state particles according to the following relations ~\cite{CLIC-Design-Report-2012,BrauJames:2007aa}:
\begin{eqnarray}\label{energy-smearing}
 \frac{\Delta  E_{jet}}{E_{jet}}  =  \frac{40 \%}{\sqrt{E_{jet}}} \oplus 2.5\% ~,~
 \frac{\Delta  E_{\ell}}{E_{\ell}}      =  \frac{15\%}{\sqrt{E_{\ell}}}\oplus 1\%
\end{eqnarray}
where $E_{jet}$ and $E_\ell$ represent the energy of the jets and leptons, respectively.
The energies are in GeV and the terms are added in quadrature. The jet energies are smeared according to a Gaussian distribution.
We smear the energies of muons similar to the electrons for simplicity. Notice that better resolutions for leptons and jet leads improve the results.
We apply the detector acceptance cuts on the transverse momenta of leptons (jets),
$p_{T}>20(30)$ GeV, and pseudorapidities, $|\eta|<2.5$. In order to have
well-isolated objects, it is required that the distances in $(\eta,\phi)$ space
between each two objects satisfies $\Delta R_{ij} = \sqrt{(\eta_{i}-\eta_{j})^{2}+(\phi_{i}-\phi_{j})^{2}} > 0.4$.
It is assumed that the presence of a high-$p_{T}$ electron or muon plays would be sufficient for triggering the signal events.

Now, the signal events are reconstructed as follows.
A full reconstruction of the W boson four-momentum ($p_{W}$) is needed to
be able to reconstruct the semi-leptonic decaying top that is the combination of the reconstructed W and $b$-jet, $M^{rec}_{W b-jet}$ .
It should yield a distribution consistent with the top-quark invariant mass, $M_{top}$.

Due to undetected neutrino which leaves no track in the detector, we have difficulties in reconstruction of the
W boson. The transverse components of the neutrino momentum ($p_{T}^{\nu}$), can be identified by the missing
transverse momentum of the events.
The longitudinal component of the momentum of the neutrino, $p_z^{\nu}$, can be found by solving the following quadratic equation:
\begin{equation}\label{p-z-nu}
 p_W^2 = ( p_\ell + p_\nu )^2 = M_W^2 \,
\end{equation}
which we put a mass constraint on W, $M_W$ = 80.4 GeV.
Solving the above quadratic equation allows us to obtain the longitudinal component of the neutrino momentum.
This equation has up to two real solutions. In the case of having two real solutions, the one with minimum absolute value is taken.
For the events with complex solution only the real part of the solution is considered as the $z$-component of the neutrino momentum.

\begin{figure*}[tbh]
\vspace{0.50cm}
\begin{tabular}{cc}
\includegraphics[width=0.45\textwidth]{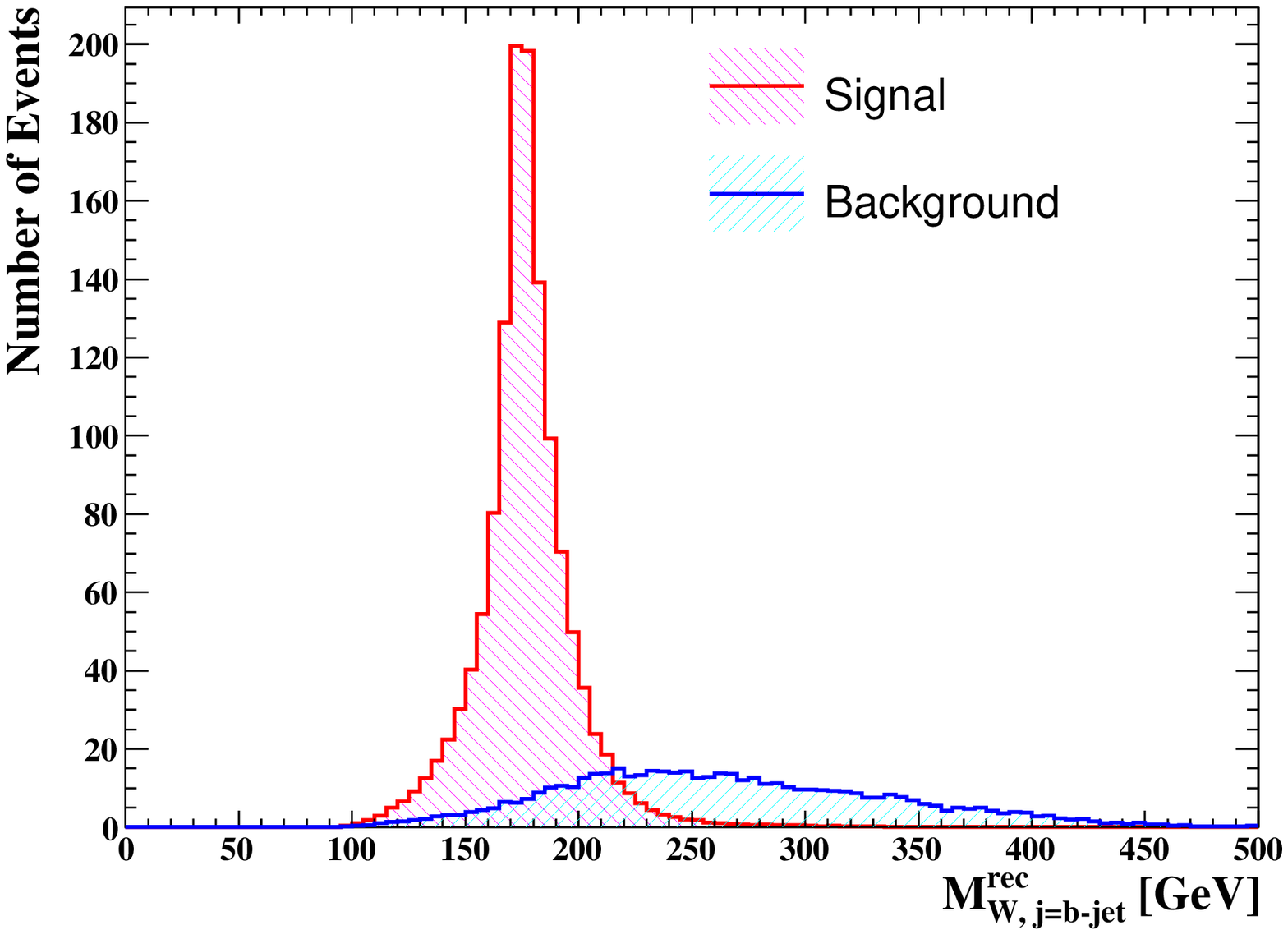} &
\includegraphics[width=0.45\textwidth]{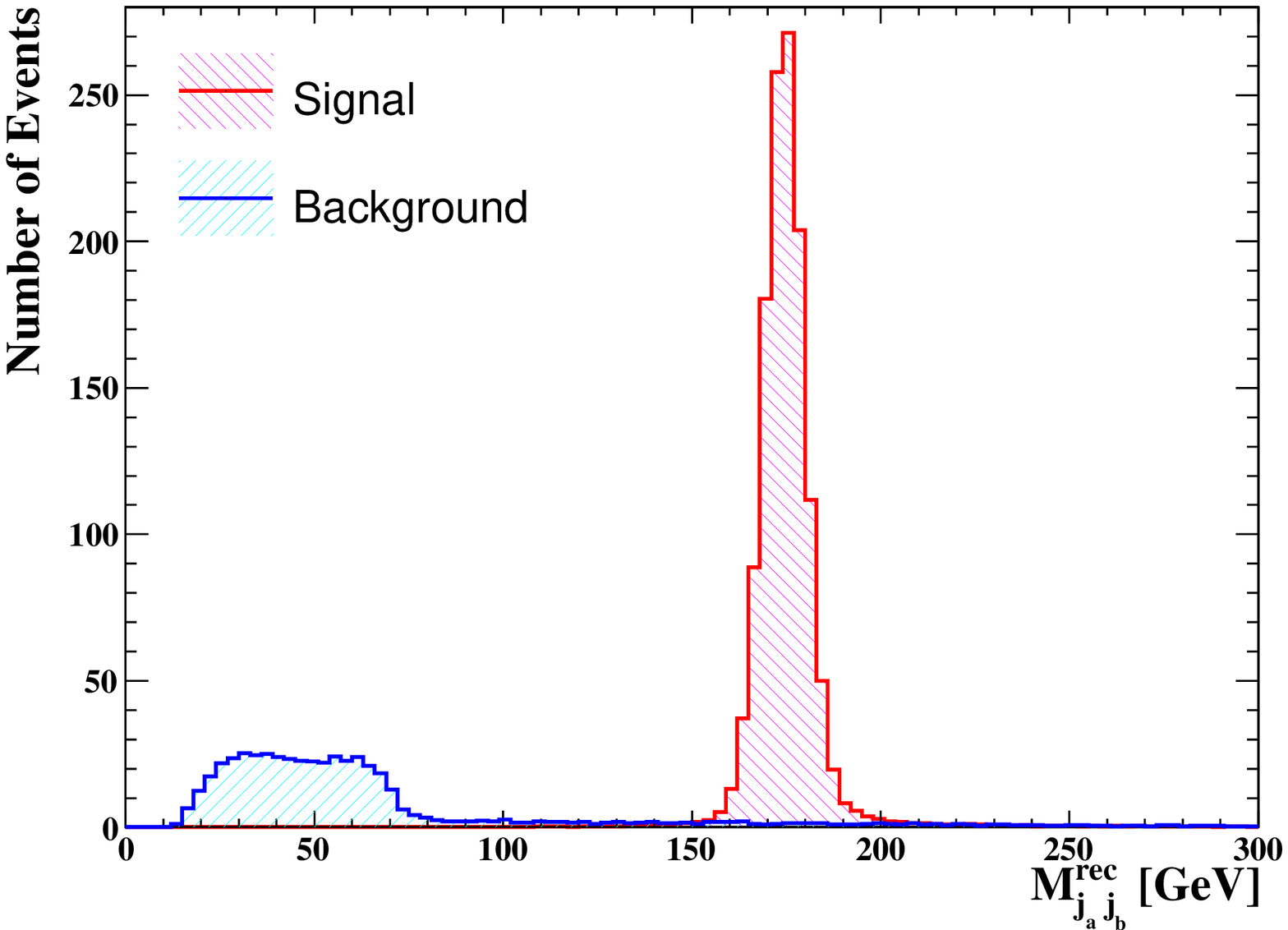}
\end{tabular}
\caption{Reconstructed top-quark mass distributions of the W-boson and b-jet $M_{W, j_c = b-jet}^{rec}$ (left),
and other two-jet $M_{j_a j_b}^{rec}$ (right) after the preselection cuts, at $\sqrt{s} = 0.5$ TeV. The number of events are
normalized to 100 fb$^{-1}$ integrated luminosity of data and for the signal $\kappa_{tqg}/\Lambda=0.02$ TeV$^{-1}$.}
\label{M-top-rec}
\end{figure*}

We assume a $b$-tagging efficiency of about 60\% for $b$-jets, 5\% for $c$-quarks and 1\% for lighter quarks
to be mis-tagged as $b$-quark jets \cite{Baer:2013cma,Abramowicz:2013tzc}.
In order to reconstruct the invariant mass of the semi-leptonic decaying top-quark, $M_{top}$,
we require a completely reconstructed W boson and a $b$-tagged jet.
The anomalously decaying top-quark will be reconstructed by combination of two other remaining jets, which are not tagged as $b$-jets.
The reconstructed invariant masses of both top quarks should have mass closest to the physical top-quark mass, $M_{top}$.
In some events, there can be more than one b-tagged jets. To make the correct combination of the jets,
the event reconstruction is completed by minimizing a $\chi^2_{abc}$ defined as:
\begin{equation}\label{chi2}
 \chi^2_{abc} = ( M_{j_a j_b}^{rec} - M_{top} )^2 + ( M_{j_c W}^{rec} - M_{top} )^2  + \Delta R^2_{j_c \ell}    \,.
\end{equation}
where $M_{j_a j_b}^{rec}$ is the reconstructed mass of the anomalously decaying top quark and
$M_{j_c W}^{rec}$ is the reconstructed mass of the top-quark decaying through SM.
$\Delta R^2_{j_c \ell}$ is the angular distance between lepton and jets.
It is expected that the jet originating from the semi-leptonic top quark decay to be to close to the charged lepton.
Various combinations of $\chi_{abc}$ are made and the one with the minimum $\chi^2$ is chosen.
The minimum value of $\chi^2$ implies that the reconstructed particles fit the requirement of coming from FCNC or SM top-quark
decay.
The reconstructed top-quark mass distribution, $M_{j_a j_b}^{rec}$ and $M_{Wj_c = b-jet}^{rec}$, for signal and corresponding background
at center-of-mass energy of $\sqrt{s} = 0.5$ TeV are shown in Fig.~\ref{M-top-rec}.
In Table \ref{cut-table} (left side), we show the number of signal and background events before and
after the kinematical cuts for an integrated luminosity of 100 fb$^{-1}$. In this table, the numbers are
presented after including the b-tagging efficiency.
%
%
%
%

\begin{figure*}[tbh]
\begin{center}
\includegraphics[width=0.40\textwidth]{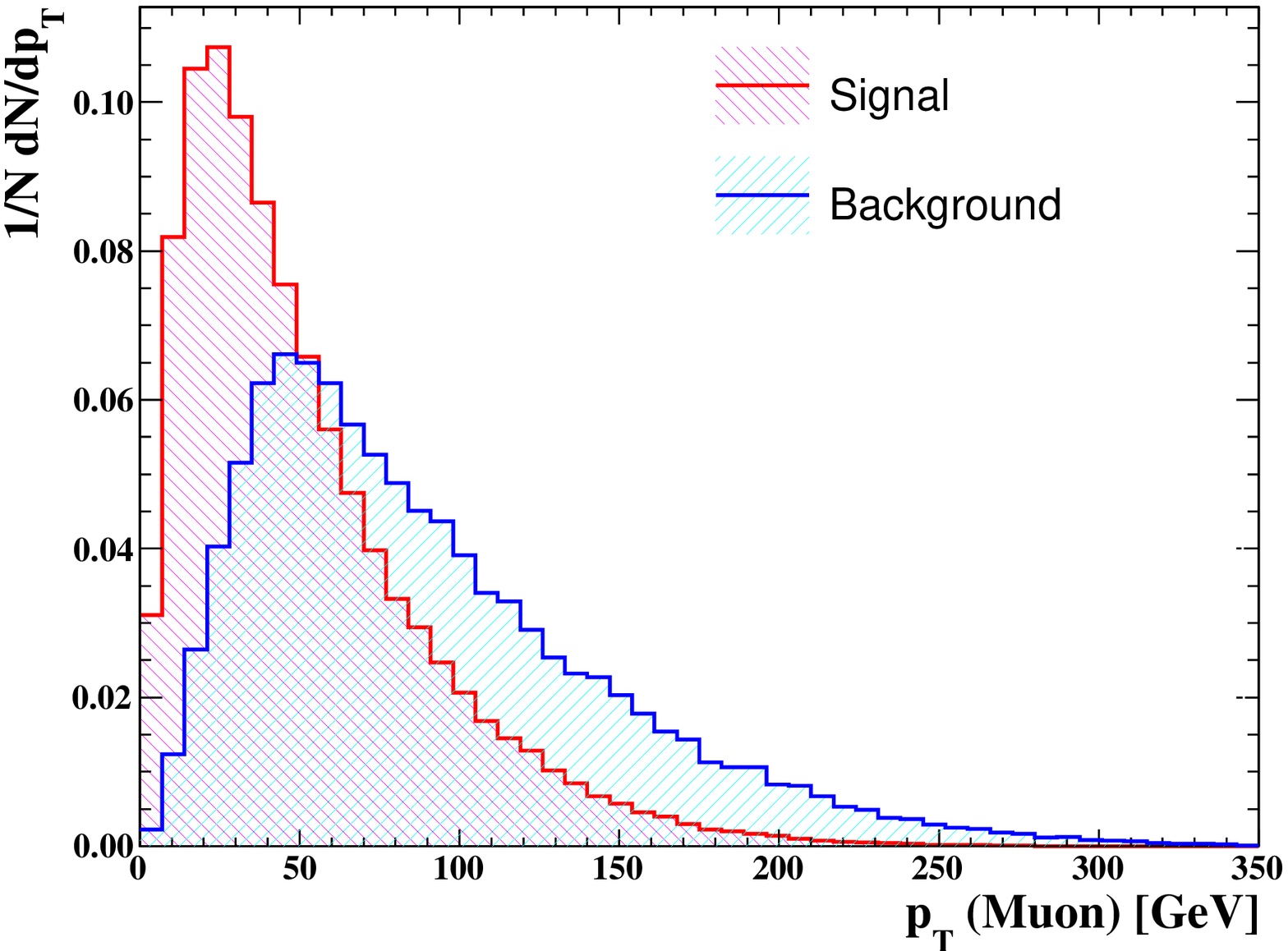}
\includegraphics[width=0.40\textwidth]{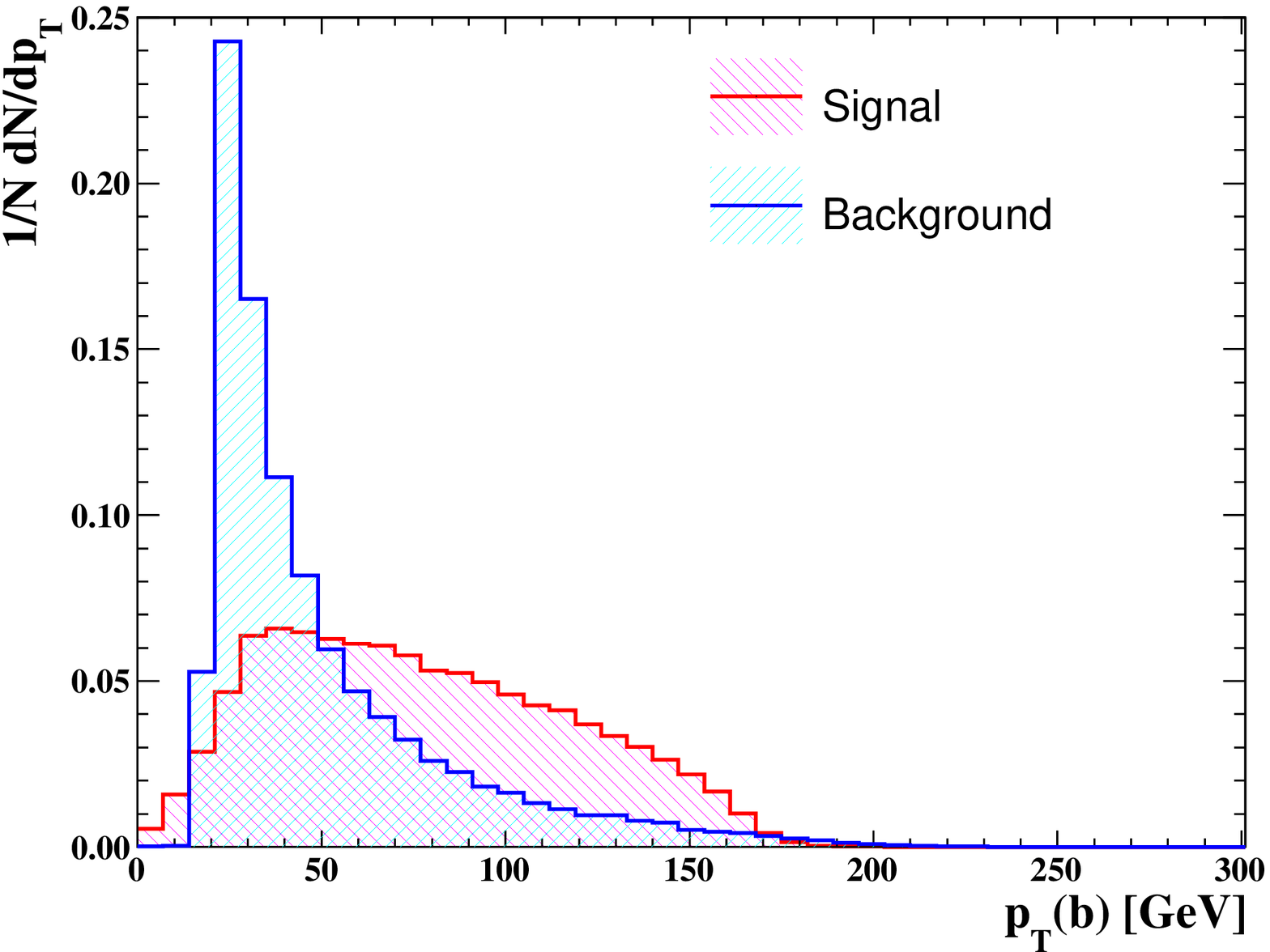} \\
(a)\hspace*{0.40\textwidth} (b) \hspace*{0.40\textwidth} \\
\includegraphics[width=0.40\textwidth]{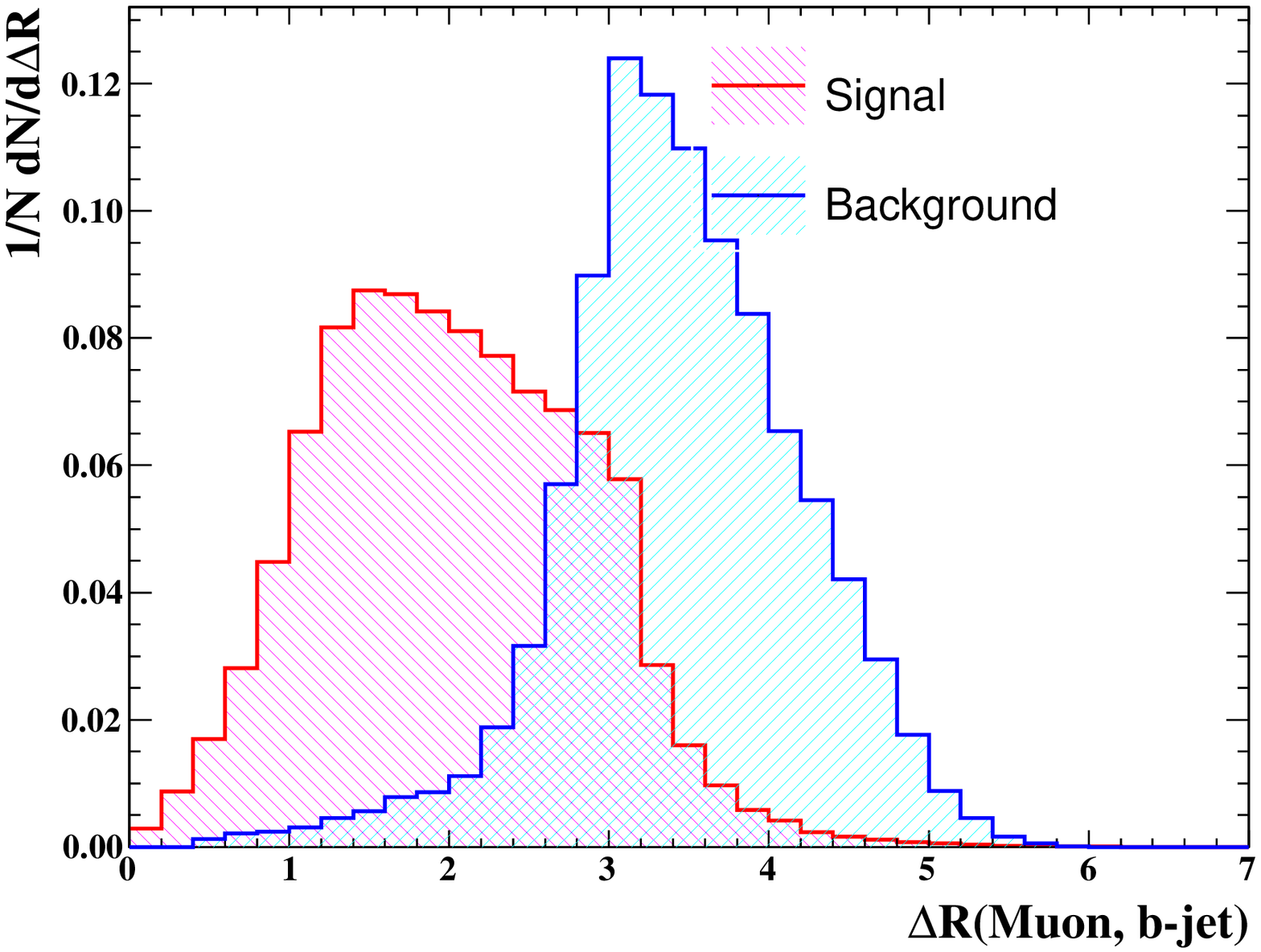}
\includegraphics[width=0.40\textwidth]{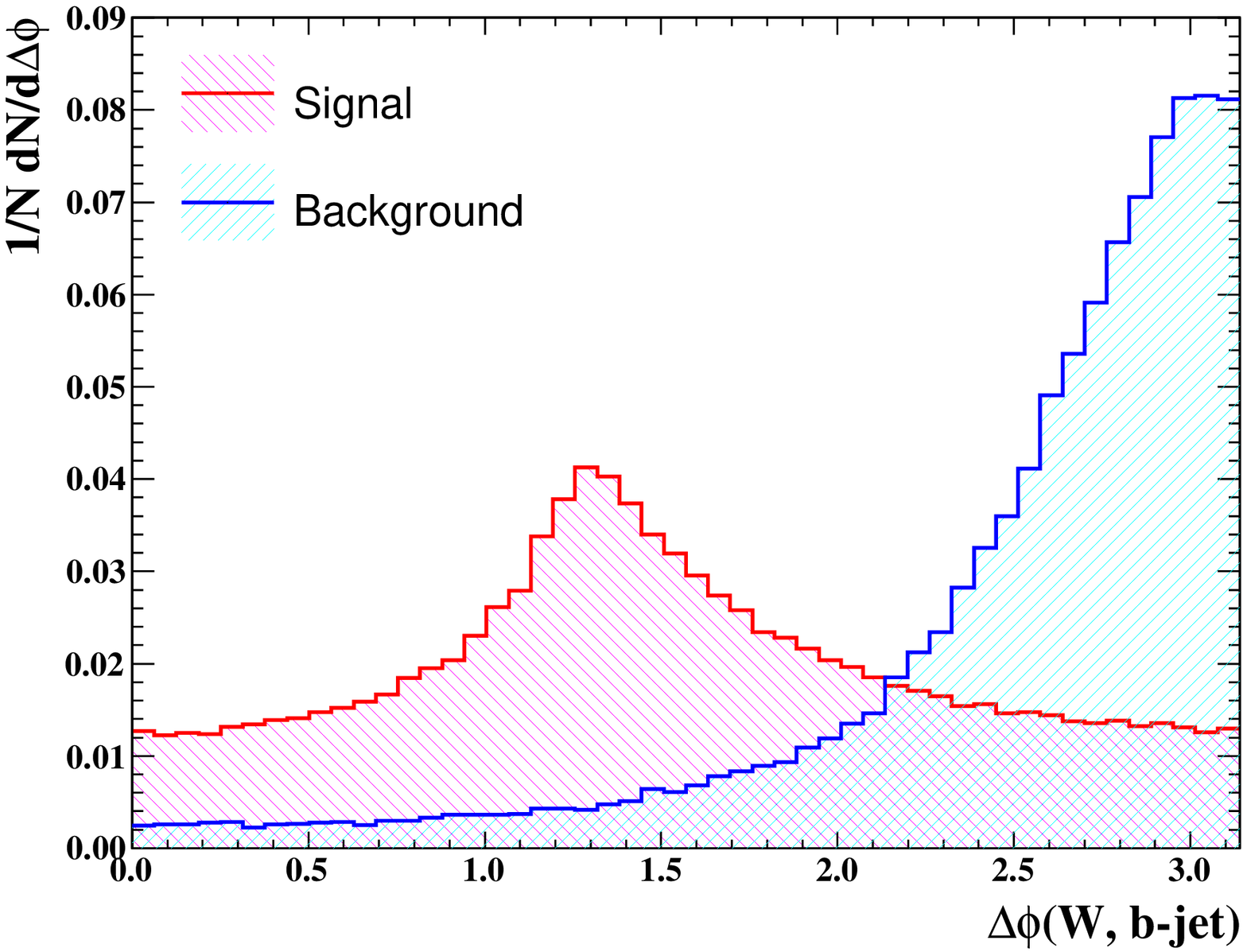} \\
(c)\hspace*{0.40\textwidth} (d) \hspace*{0.40\textwidth} \\
\caption{The kinematic distributions of four significance variables used as inputs to BDT in addition to the $W$ boson and top
mass distributions.
(a) The transverse momentum of the charged lepton $p_T (\ell)$ and
(b) transverse momentum of the $b$-jet $p_T (b)$.
(c) The angular separation between charged lepton and $b$-jet $\Delta R_{\ell, b-jet}$ and
(d) the difference of the azimuthal angle between the reconstructed W boson and the $b$-jet $|\Delta \phi _{W, b-jet}|$.}
\label{Dicrimination-variables}
\end{center}
\end{figure*}

Certainly, a detailed background study is essential in order to separate the
signal from the background events. We use TMVA~\cite{Therhaag:2010zz,TMVA2} as a toolkit
for multivariate analysis to separate signal from background.
Indeed, a multivariate analysis technique is necessary because a single variable doesn't have sufficient discrimination power
to separate signal from background events. Among the multivariate analysis techniques are used to separate the signal from the
backgrounds boosted decision trees (BDT) is chosen ~\cite{Roe:2004na,oai:arXiv.org:physics/0508045,ANN}.
We choose the variables which have the most possible separation power between the signal from background events for the
BDT input.
The kinematical variables are selected as input to the BDT are as follow:
The reconstructed top quarks masses, the transverse momenta of the $b$-jet $p_T (b)$ and charged lepton $p_T (\ell)$,
the difference of the azimuthal angle between the reconstructed W boson and the $b$-jet $|\Delta \phi _{W, b-jet}|$,
the pseudorapidity distribution of $b$-jet $|\eta_{b-jet}|$, the invariant mass of the reconstructed W boson $M^{rec}_{\ell \nu}$
, and finally the angular separation between charged lepton and $b$-jet $\Delta R_{ \ell, b-jet}$.
The variable list can be extended and more variables could be given in the BDT input for better discrimination
between signal and backgrounds.
The kinematic distributions of some used variables which has the most discrimination power are presented
in Fig.~\ref{Dicrimination-variables} before applying the acceptance cuts. These distributions are normalized to unity.

%
%
%
%

These variables  are given to the BDT and the multivariate analysis is performed to achieve the best separation
between signal and backgrounds and enhance signal significance.
The test and training processes are done using a mixture of 50\% of signal and 50\% of background events.
Due to the sensitivity of the BDT classifier to the statistical fluctuation of the training data sample,
we use Adaptive Boosting algorithm to increase the performance.
In order to avoid overtraining and to improve the quality of the analysis,
the BDT built-in options such as Cost Complexity pruning methods are implemented during the training process.
The goal is to find the best cuts that enhance the signal and reduce the background.
Obtaining best cuts is generally done by finding the maximum value of the statistical significance, $\frac{n_{s}}{\sqrt{n_{s} + n_{b}}}$, where
$n_{s}$ is the number of signal events and $n_{b}$ is the number of background event.
By choosing the optimum cuts on the BDT output spectrum, we determine the number of selected signal and background events
that provide the best signal significance. The results will be discussed in Section~\ref{results}.

%
%
\section{Fully-hadronic decays of the $t \overline{t}$}\label{Fully-hadronic}

In pervious section, we discussed the signal and background where one of the top
quarks decays through SM decay mode $t \rightarrow b W \rightarrow \ell \nu b$ and the other one is considered to decay through FCNC into $t \rightarrow qg$.
In this section, we consider FCNC decay of both top quarks
where the final state consists four-jets at  the center of mass energies of $\sqrt{s} = 0.5,~ 1$ and $1.5$ TeV.
Specifically, the final state is characterized by the $t \bar{t}$ events with both top quarks decay into a gluon and a light quark, $t \rightarrow q g$.

\begin{figure}[tbh]
\vspace{0.75cm}
\centerline{\includegraphics[width=0.50\textwidth]{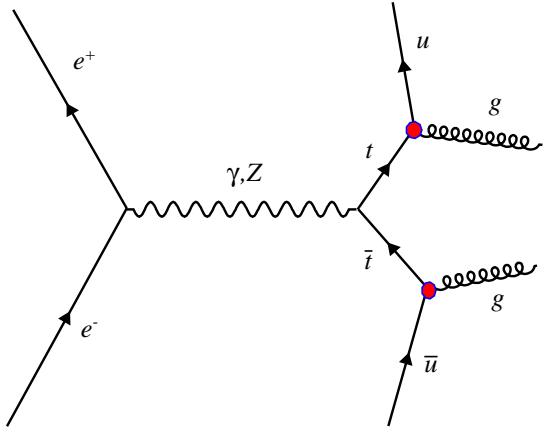}}
\caption{The representative Feynman diagram for the signal process in the hadronic channel.}
\label{feynhad}
\end{figure}

The representative Feynman diagram for the signal process in the full-hadronic channel is depicted in Fig.\ref{feynhad}.

%
%

It is worth mentioning that hadron colliders may not be a good area to study this fully hadronic process due to the
extremely large QCD background contributions.
The linear electron-positron colliders such as CLIC or ILC have a clean environment,
consequently these fully hadronic final states can be probed at the CLIC or ILC easier than the hadron colliders.

The method of the channel is similar to semi-leptonic one that presented in the previous section.
The MC generation of the signal sample is generated with { \sc CompHEP} and the complete set of four-jets background is done
using { \sc MadGraph} 5.

\begin{figure*}[tbh]
\vspace{0.50cm}
\begin{tabular}{cc}
\includegraphics[width=0.45\textwidth]{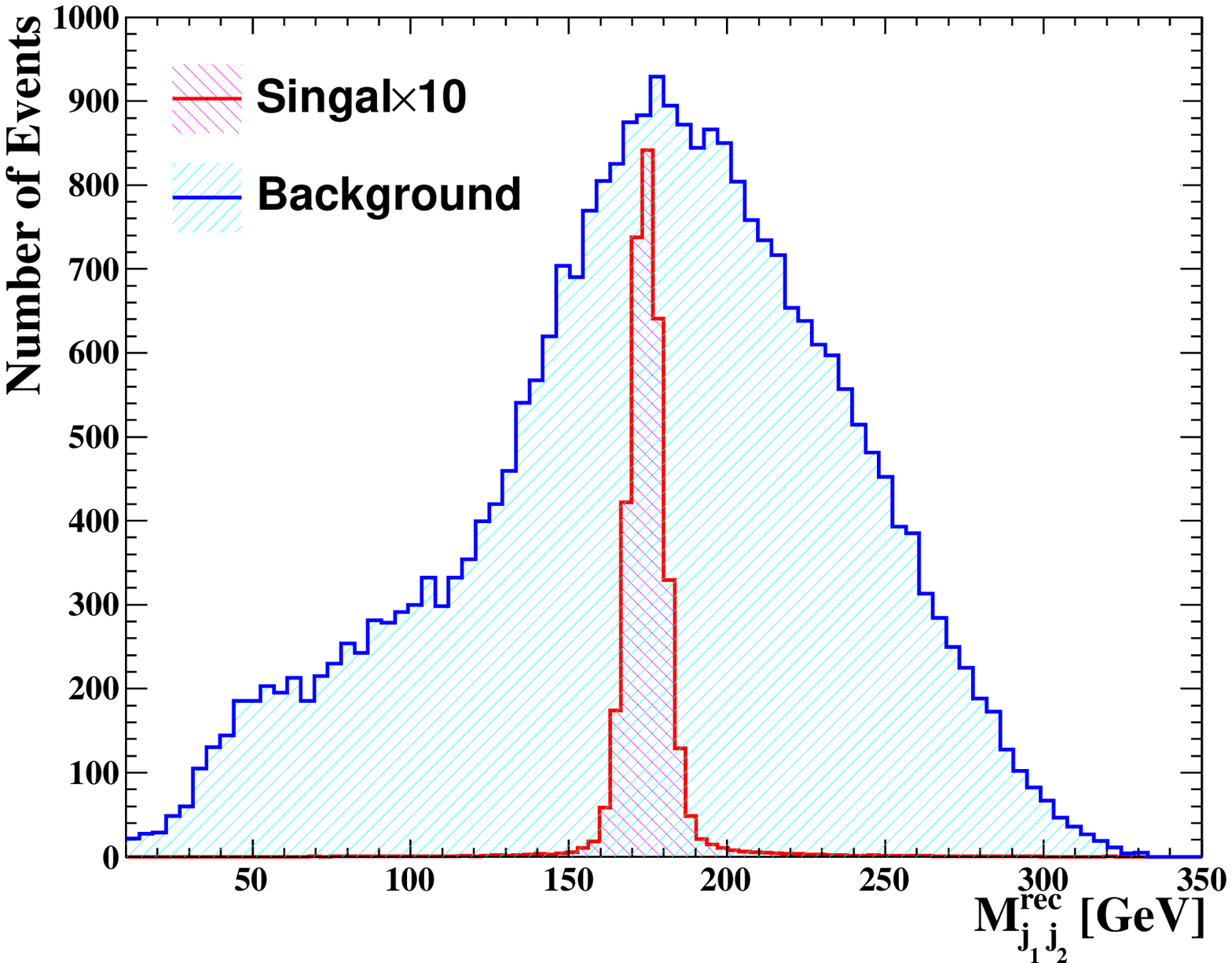} &
\includegraphics[width=0.45\textwidth]{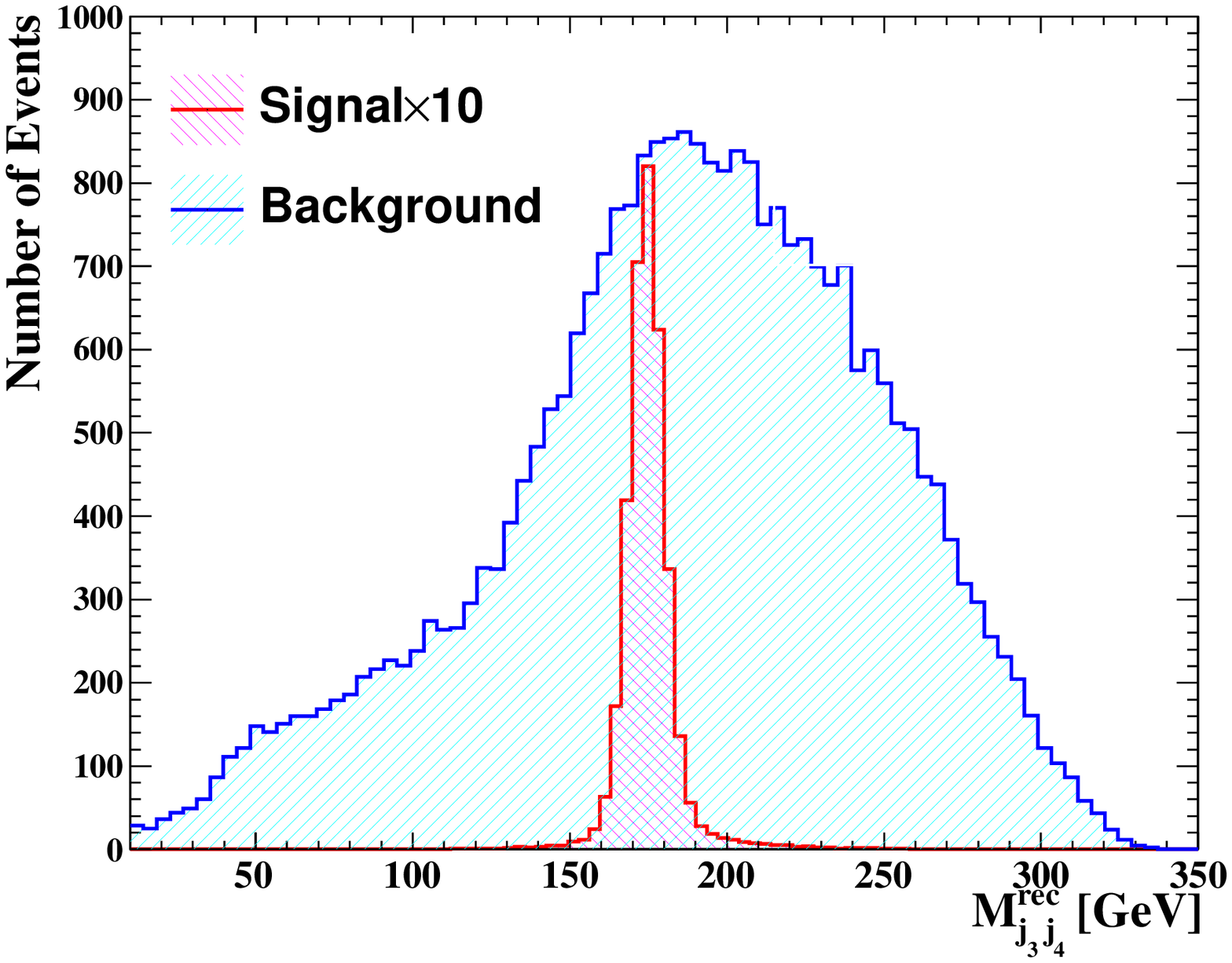}
\end{tabular}
\caption{Reconstructed top-quark mass distributions for the anomalous decay of both top quarks into a light-quark and a gluon
at $\sqrt{s} = 0.5$ TeV. The distributions are normalized to 100 fb$^{-1}$ integrated luminosity of data  and for the signal $\kappa_{tqg}/\Lambda=0.02$ TeV$^{-1}$. The number of signal events are multiplied by a factor 10.}
\label{M-top-full-hadronic}
\end{figure*}

The same as the semi-leptonic case, to account for the resolution of the detectors a Gaussian energy smearing is
performed on the final states jets. The jets are required to have transverse momenta greater than 30 GeV within
the pseudorapidity  acceptance range of $|\eta| < 2.5$. It is also required that  $\Delta R_{ij} = \sqrt{(\eta_{i}-\eta_{j})^{2}+(\phi_{i}-\phi_{j})^{2}}>0.4$.
The number of events before and after the kinematical cuts are shown in the right side of Table~\ref{cut-table} for an
integrated luminosity of 100 fb$^{-1}$.
The reconstructed top-quark mass distributions for signal and the corresponding background
at center-of-mass energy of $\sqrt{s} = 0.5$ TeV are shown in Fig.~\ref{M-top-full-hadronic}
for an integrated luminosity of 100 fb$^{-1}$. The number of signal events in these figures
have been multiplied by a factor of 10.

Again we use the Boosted Decision Tree (BDT) classifier of the TMVA package for discriminating signal from background events.
For the BDT algorithm, the simulated events of the signal and background are split up in two similar samples for the training and test processes.

The input kinematical variables to the BDT are the
reconstructed top-quark masses $M^{rec}_{top}$, the transverse momentum
of the highest $P_{T}$ jet, the corresponding pseudorapidity distribution $|\eta_{j}|$ of the highest $P_{T}$
jet, the angular separation $\Delta R_{j_{a} j_{b}}$ between the two jets, and the scalar transverse energy, $H_{T}$.
The opening angles $\Delta \phi_{j_{a} j_{b}}$ between the directions of the final state jets is correlated with the mentioned variables so
we neglected them. For the fully hadronic top-quark reconstruction, we take the pair of jets which have an invariant mass closest
to the nominal top-quark mass as well as having smaller angular
distance $\Delta R_{j_{a} j_{b}}$.  In the analysis, an angular
resolution of around 100 mrad is assumed due to the expected high
granularity design of the calorimeters of the future electron-positron
collider \cite{resolution}. 

%
%

In summary, in this section we concentrated on the channel of top pair production which both top quarks decay anomalously into two jets.
After a rough detector simulation and applying the acceptance cuts, optimum kinematical variables are found and given to the BDT for discriminating between
signal and backgrounds. In the next section, the limit on the branching fractions are given.

%
\begin{table}[tbh]
\caption{The number of events before and after the kinematical cuts for signal and background at a center-of-mass energy
of 0.5 TeV and with 100 fb$^{-1}$ integrated luminosity of data for semileptonic and full hadronic (both top decay anomalously). The b-tagging
efficiencies have been included and for the signal we have set $\kappa_{tqg}/\Lambda = 0.02$ TeV$^{-1}$.}
\begin{center}
\begin{tabular}{c|cc|cc}
\hline
     Decay mode          &  ~~~~~~~~~~~~~~~~~Semi-leptonic   &    &  ~~~~~~~~~~~~~~~~~Full-hadronic  &     \\ \hline
                                  &  Before cuts   & After cuts   &  Before cuts  &   After cuts  \\ \hline
           Signal             &     2146.5       & 1297.4        &  480.0           &   354.0 \\
           Background    &     1743.2       &   500.0        &   58905.1      &   29211.3\\
  \hline
\end{tabular}\label{cut-table}
\end{center}
\end{table}

%
%
\section{ Results }\label{results}
With assuming of observation no signal events after performing the experiment or
in another words, if the number of observed events are equal to the number
of expected background events, we proceed to set $95\%$ C.L. upper limit on the signal cross section.
Limits on the cross section of signal is calculated with a CLs approach \cite{cls}.
The RooStats \cite{roostat} program is used for statistical data analysis  for the numerical evaluation of
the CLs limits. The program returns the $95\%$ C.L. upper limit on the signal cross section times
branching ratios of the top quarks decays.

\begin{figure}[tbh]
\centerline{\includegraphics[width=0.48\textwidth]{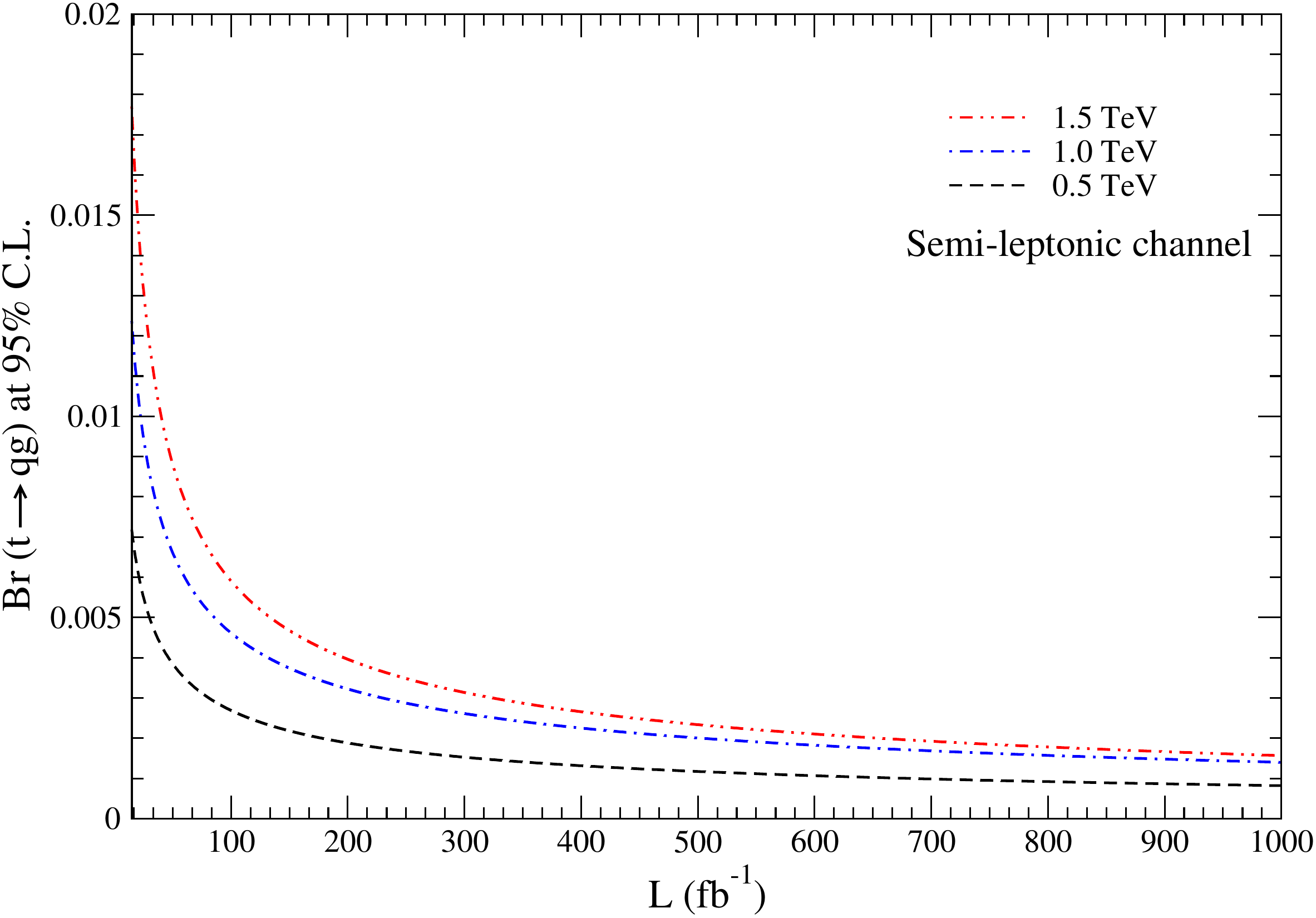}}
\caption{ The $95\%$ C.L. upper limits for $Br(t\rightarrow qg )$ as a function of integrated
luminosity for $ \sqrt{s} = 0.5 $, $1$ and $1.5$ TeV for the semi-leptonic analysis.}
\label{Br-luminosity-leptonic}
\end{figure}

\begin{figure}[tbh]
\centerline{\includegraphics[width=0.48\textwidth]{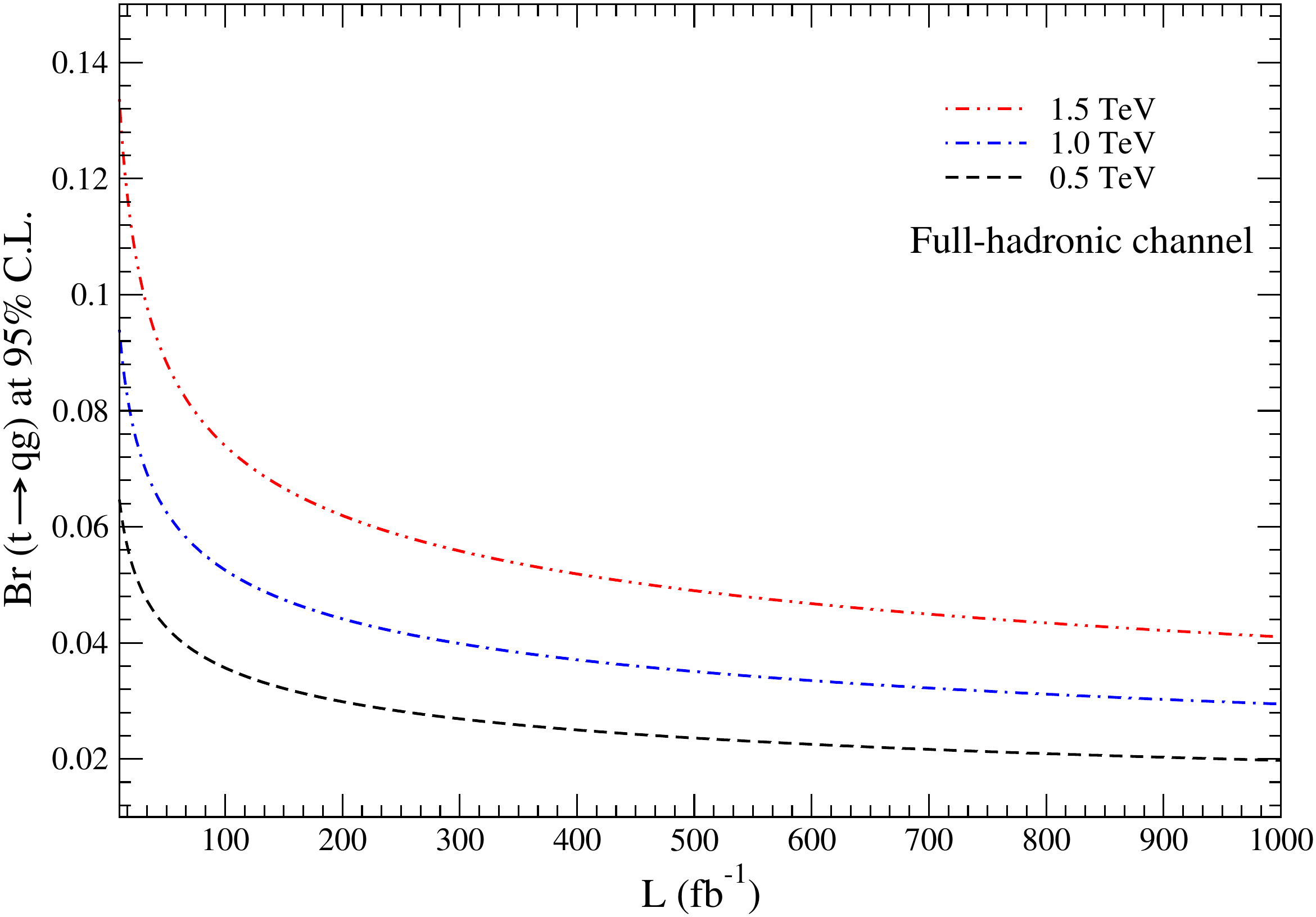}}
\caption{The $95\%$ C.L. upper limits for $Br(t\rightarrow qg )$ as a function of integrated luminosity for
$ \sqrt{s} = 0.5 $, $1$ and $1.5$ TeV for the fully-hadronic analysis. }
\label{Br-luminosity-hadronic}
\end{figure}

The sensitivity of the branching fractions as a function of the integrated luminosity
for the future electron-positron collider at different center-of-mass energies are shown in Fig.~\ref{Br-luminosity-leptonic}
and Fig.~\ref{Br-luminosity-hadronic} for the semi-leptonic and fully-hadronic analyses, respectively.
As it can be seen from the figure, higher integrated luminosities lead to  better bounds on the branching ratio
up to around 500 fb$^{-1}$.
The limits at the center of mass energy of 0.5 TeV is better than the ones at 1 and 1.5 TeV that is because
of the larger cross sections at smaller energies.
Comparing the semi-leptonic channel with the full-hadronic one, better sensitivity is achieved
in the semi-leptonic channel. It is again due to the fact that in the full-hadronic channel statistics
is poor with respect to the semi-leptonic one.
The upper limits on the $Br(t\rightarrow qg )$ at 95\% C.L. with 500 fb$^{-1}$ are 0.00117
and 0.0236 for the semi-leptonic and full-hadronic channels, respectively.
It is interesting to mention here that the dependence of the expected upper limit on the 
integrated luminosity becomes weaker at luminosities larger than 500 fb$^{-1}$.

Now, the sensitivity of the results on the detector performance is discussed. 
In this analysis, almost all sub-detectors are involved to identify and reconstruct 
leptons, jets, b-jets and missing energy. Precise reconstruction of secondary vertex
for an efficient b-tagging is necessary in this analysis to suppress the backgrounds
and obtain a pure signal sample. The variation of b-tagging efficiency in this analysis
by $10\%$ leads to approximately $4\%$ change in the expected upper limit on the branching 
ratio. The resolution in measurement of the jet and lepton energies are less important than 
b-jet identification. Varying the resolution in jet and lepton energy measurement by
$10\%$ and $5\%$ (Eq.\ref{energy-smearing}) leads to change the upper limit on the branching 
fraction by less than $1\%$.

In this analysis, we have calculated the cross section for the energy at the 
vertex of electron-positron. Therefore, further effects such as the
initial state radiation (ISR) and the luminosity spectrum (LS) of the collider need to 
be considered. Both the initial state radiation 
and the luminosity spectrum lead to reduce the cross
section.  We calculate the effect of ISR to the cross section
at the center-of-mass energy of 500 GeV. 
The signal cross section decreases by a round $2\%$ which leads to
loose the expected upper limit on the branching ratio from 0.00117 
to 0.00118 with 500 fb$^{-1}$.

To have a more realistic analysis,  the effects of the systematic uncertainties
should be estimated.  The uncertainties can arise from jet energy scale, lepton energy, 
lepton reconstruction identification efficiencies, b-tagging efficiency and
uncertainties on the masses of the top quark and W-boson. 
We vary the b-tagging efficiency by $\pm 5\%$. This 
leads to change the expected upper limit by $1.5\%$.  
To estimate the uncertainty from jet energy scale, we vary the energy of each 
jet by $2\%$ and recalculate the limit. It results to a change of $0.5\%$ on the expected upper limit.
The uncertainties on the top quark and W boson masses are calculated as follows: We generate new signal
samples with varied top mass ($\pm$ 1 GeV) and W boson mass ($\pm 50$ MeV) and re-do the analysis. 
This leads to a change of $0.05\%$ on the expected upper limit on the branching fraction
of $t\rightarrow qg$.

%
%
%
%
\vspace{1cm}
%
%
%
%

\subsection{Comparison with the LHC Results}

So far, we have examined the future $e^{-}e^{+}$ collider potential to probe the anomalous $tqg$ in the decay of the
top quark in top pair production. At hadron colliders, the anomalous $tqg$ couplings can be probed
either in top production or top decay. The best limits have been obtained in the production processes.
There are different production channels to search for the
anomalous $t q g$: (1) direct top quark production ($2 \rightarrow 1$ process), (2) Single top quark production ($2\rightarrow 2$),
(3) double top pair ($tt, \bar{t}\bar{t}$) production and (4) top plus vector boson production ($t V$) \cite{beneke}.
 Currently,  the strongest experimental limits on the $t q g$ branching fractions come from
the direct top production ($2 \rightarrow 1$ process) at the LHC by the
ATLAS Collaboration, $ Br(t \rightarrow ug) < 3.1 \times 10^{-5}$ and
$Br(t \rightarrow cg) < 1.6 \times 10^{-4}$ at a center-of-mass energy of $\sqrt{s} = 8$ TeV
corresponding to an integrated luminosity of ${\cal L}_{int} = 14.2$ fb$^{-1}$~\cite{TheATLAScollaboration:2013vha}.

In \cite{Han:1996ce}, the anomalous $tqg$ couplings have been probed in top decay at the Tevatron.
The obtained upper limit
on the branching ratio is $5(2.7)\times 10^{-3}$ with 10 (30) fb$^{-1}$ of data. These limits
are weaker in comparison with the limits that can be obtained from the production processes.

The future LHC bounds at 14 TeV center-of-energy using 100 fb$^{-1}$ of data  using various
processes are compared with the ones obtained in this work are compared in Table \ref{compare}.
As it can be seen, among the all processes the $2\rightarrow 1$ process provides the strongest
limit ($10^{-6}$).
The limits that we have obtained in this study for $e^{-}e^{+}$ collider are comparable
with the ones that come out of the same-sign top ($tt, \bar{t}\bar{t}$) production at the LHC
and the ones from top decays in top pair events at the Tevatron.
%
\begin{table}[tbh]
\caption{The $95\%$ C.L. upper limit on the $Br(t \rightarrow qg)$ with the LHC (8, 14 TeV) and $e^{-}e^{+}$ (0.5 TeV) based on
100 fb$^{-1}$ integrated luminosity of data. The results of LHC8 is corresponding to 14.2 fb$^{-1}$.}
\begin{center}
\begin{tabular}{c|c|cccc|c}
\hline
  Collider     & LHC8 (14.2 fb$^{-1}$)  &  LHC14  (100 fb$^{-1}$)  &                   &                          &  &  $e^{-}e^{+}$ (100 fb$^{-1}$)  \\ \hline \hline
  Process       & $2 \rightarrow 1$  &  $2 \rightarrow 1$  &  $2 \rightarrow 2$ &  $tV$                    &  $tt, \bar{t}\bar{t}$  &   $t\bar{t}$  \\ \hline
  Upper limit   & $3.1 \times 10^{-5}$ &      $10^{-6}$      &   $10^{-5}$         &   $10^{-5}$          &  $10^{-3}$     &             $10^{-3}$   \\
  \hline
\end{tabular}\label{compare}
\end{center}
\end{table}
%

%
%
\section{Summary and conclusions} \label{Conclusions}

In this paper, we have studied the signals of top quark flavor-changing neutral current in
the vertex of $tqg$, where $q=u$ and $c$, at a future electron-positron collider.
This study has been done by looking at the top pair production and at three different
center of mass energies of 0.5, 1 and 1.5 TeV in the top quarks decays.
We have investigated two possible cases: first one is the case that one top quark decays anomalously to $q+g$
and another one follows SM decay to a W boson and a b-quark and W boson decays leptonically
($e^{-}e^{+} \rightarrow t\bar{t} \rightarrow q g \ell^{+} \nu_{\ell}b$). Second is that both top quarks
decay anomalously through FCNC decay mode ($e^{-}e^{+} \rightarrow t \bar{t} \rightarrow q\bar{q}gg$).
Using the Boosted Decision Tree (BDT) technique, we discriminate between signal and backgrounds.
Then the CLs approach has been utilized to set upper limits on the branching ratio.
The 95\% C.L. upper limit on the branching ratio using 500 fb$^{-1}$ of data at the center
of mass energy of 0.5 TeV is 0.00117 (0.0236) in semi-leptonic (full-hadronic) channel.
It is shown that the limit is improved with the integrated luminosity up to around 500 fb$^{-1}$
and the dependence of the expected upper limit on the integrated luminosity becomes weaker at
luminosities larger than 500 fb$^{-1}$.

We have found that sensitivity to the anomalous couplings decreases with increasing
the center of energy of the collisions simply due to the decrease in the signal cross
section with growing the center of mass energy.
The expected bounds are comparable with the ones that is obtained from the double top production  at the LHC ($tt,\bar{t}\bar{t}$)
and from the anomalous top decay in top pair events at Tevatron.

%
%
\section*{Acknowledgments}

We are grateful to R. Goldouzian for valuable discussions.
We acknowledge financial support of the School of Particles and Accelerators,
Institute for Research in Fundamental Sciences (IPM).
%

%
%
\section*{Conflict of Interests}

The authors declare that there is no conflict of interests regarding the publication of this paper.

%
%
%


\end{document}